\documentclass[conference]{IEEEtran}

\usepackage{cite}
\usepackage{amsmath,amssymb,amsfonts}

\usepackage{algorithm}
\usepackage{algpseudocode}

\usepackage{graphicx}
\usepackage{textcomp}
\usepackage{xcolor}

\usepackage{multirow}
\usepackage{svg}

\usepackage{subcaption}

\usepackage{array}
\usepackage{adjustbox}

\def\BibTeX{{\rm B\kern-.05em{\sc i\kern-.025em b}\kern-.08em
    T\kern-.1667em\lower.7ex\hbox{E}\kern-.125emX}}
\begin{document}

\title{Deterministic Sort-Free Candidate Pruning for Scalable MIMO Box Decoding
}

\author{
\IEEEauthorblockN{Shengchun Yang, Amit Sravan Bora, Emil Matus, Gerhard Fettweis}
\IEEEauthorblockA{Vodafone Chair Mobile Communications Systems, Technische Universität Dresden, 01062 Dresden, Germany \\
Email: \{shengchun.yang, amit.bora, emil.matus, gerhard.fettweis\}@tu-dresden.de}
}
\maketitle

\begin{abstract}
Box Decoding is a sort-free tree-search MIMO detector whose complexity is independent of the QAM order, achieved by searching a fixed candidate “box” around a zero-forcing (ZF) estimate. However, without pruning, the number of visited nodes grows exponentially with the MIMO dimension, limiting scalability. This work proposes two deterministic, low-complexity, sort-free pruning strategies to control node growth. By exploiting the geometric symmetry of the QAM grid and the relative displacement between the ZF estimate and nearby constellation points, the proposed methods eliminate unnecessary metric evaluations while preserving QAM-order independence. The resulting detector achieves substantial complexity reduction with negligible error-rate degradation and enables fully parallel, hardware-efficient implementations for large-scale MIMO and higher-order QAM systems.
\end{abstract}

\begin{IEEEkeywords}
MIMO detection, Box Decoding, Candidate pruning, $\boldsymbol{K}$-Best, Sort-free, Large-scale MIMO.
\end{IEEEkeywords}


\section{Introduction}
\noindent
MIMO detection has been extensively studied in communication systems, yet managing the performance–complexity trade-off remains challenging as antenna dimensions scale in 5G and beyond \cite{MIMO_detection_Hanzo, MIMO_Larsson}. Optimal nonlinear detectors such as maximum likelihood (ML) and maximum a posteriori (MAP) provide the best error-rate performance, but their computational cost grows exponentially with modulation order and MIMO dimension \cite{MIMO_Larsson}. At the other extreme, linear receivers, including Zero Forcing (ZF) and linear minimum mean square error (LMMSE), are attractive for hardware implementation due to their simple processing flow, but they suffer from substantial performance degradation, particularly in ill-conditioned channels \cite{MIMO_Larsson}.\par

To bridge this gap, tree-search detectors have been widely investigated. Depth-first sphere decoding (SD) \cite{sphere_viterbo} and breadth-first $K-$Best search \cite{K-best_JSAC} offer practical trade-offs between complexity and performance. However, $K-$Best typically requires sorting, which is inherently hardware-unfriendly, involving wide comparator networks and deep critical paths \cite{k_best_sort_free,KBest_VLSI}. Existing work has explored faster sorter architectures \cite{KBest_Sorter_2017,Sorter_2023} and techniques that reduce the number of candidates before sorting \cite{cha_perm2025,New_SD_2025}. Sort-free variants have also been reported, including partial-expansion approaches \cite{sort_free_Globecom} and Distributed $K-$Best (DKB) that rely on iterative selection via multi-way merge networks \cite{DKB_conf,k_best_lattice_reduction,multiway}, with further reductions in iteration count using discretized path-metric search \cite{sort_free_2024}. Nevertheless, these $K-$Best-based approaches still rely on enumeration-based expansion, keeping the overall complexity tied to the constellation size \cite{k_best_sort_free,KBest_VLSI}.\par

Recently, Box Decoding was proposed as a sort-free alternative to $K-$Best that avoids constellation-wise enumeration \cite{box}. At each detection layer, it selects a fixed-size candidate grid (``box'') around a ZF estimate, resulting in a complexity that is independent of the QAM order. However, it was originally demonstrated only for $2 \times 2$ MIMO, as the number of visited nodes grows exponentially with the MIMO size. $K$-Box \cite{KBox} mitigates this growth by reintroducing sorting, but the associated overhead remains hardware-unfriendly. In our recent work \cite{Box_Prob}, we proposed a sort-free strategy that prunes unlikely branches using a statistically derived threshold. Although this approach remains sort-free and QAM-order-independent, the number of visited nodes varies with SNR, which limits full parallelization in hardware.\par

This paper addresses both the \textit{scalability} and \textit{hardware parallelization} challenges of Box Decoding. The main contributions are summarized as follows:
\begin{enumerate}
    \item  We propose two deterministic, low-complexity pruning strategies—\textit{Single-Step Candidate Pruning (SCP) and Iterative Candidate Pruning (ICP)}, and  their hybrid extension, to scale Box Decoding to large MIMO.
    \item The proposed pruning rules exploit the \textit{geometric} symmetry of the QAM grid and the \textit{relative position} between the ZF estimate and constellation points, thereby frequently bypassing full Euclidean-distance computations.
    \item The resulting detector remains \textit{sort-free} and \textit{QAM-order-independent}, and naturally maps to \textit{parallel hardware architectures}. Simulation results demonstrate substantial complexity savings with negligible bit error rate (BER) degradation.\par

\end{enumerate}


\textbf{Notations:} $a$, $\boldsymbol{a}$, and $\boldsymbol{\mathrm{A}}$ denote a scalar, a vector, and a matrix, respectively. $\boldsymbol{\mathrm{A}}^H$ represents the Hermitian of $\boldsymbol{\mathrm{A}}$, whereas $|\boldsymbol{\mathrm{a}}|^2$ and $\lfloor . \rfloor$ denote the norm-squared and the floor operation, respectively. Finally, symbols $\mathbb{C}$, $\mathfrak{R}$, and $\mathfrak{I}$ denote the complex field, and its real and imaginary parts, respectively.

%


\section{System Model}
\subsection{Tree-Based MIMO Detection}

Consider a $N\times N$ MIMO system with transmitted symbol vector $\mathbf{x}\in\mathcal{A}^N$, received symbol vector $\mathbf{y}\in\mathbb{C}^N$ and channel matrix $\mathbf{H}\in\mathbb{C}^{N\times N}$, where $\mathcal A$ denotes the QAM set with size $A = |\mathcal A|$. To enable tree-search detection, applying the QR decomposition $\mathbf{H}=\mathbf{Q}\mathbf{R}$ (unitary $\mathbf{Q}$, upper-triangular $\mathbf{R}$) yields
\begin{equation}
\tilde{\mathbf{y}}\triangleq \mathbf{Q}^H\mathbf{y}=\mathbf{R}\mathbf{x}+\tilde{\mathbf{w}},
\end{equation}
where $\tilde{\mathbf{w}}=\mathbf{Q}^H\mathbf{w}$ remains white due to $\boldsymbol{\mathrm{Q}}$ being unitary. The search proceeds from layer $i=N$ to $i=1$ (last to first row of $\mathbf{R}$), where the partial Euclidean distance (PED) in each layer is calculated recursively as 
\begin{equation}
d_i=d_{i+1}+\left|\tilde{y}_i-r_{i,i}\hat{x}_i-\sum_{j=i+1}^{N}r_{i,j}\hat{x}_j\right|^2,
\label{PEDa}
\end{equation}
with $d_N=|\tilde{y}_N-r_{N,N}\hat{x}_N|^2$ at layer $i=N$. 

\begin{figure}[t]
        \centering 
        \includegraphics[width=0.3\textwidth]{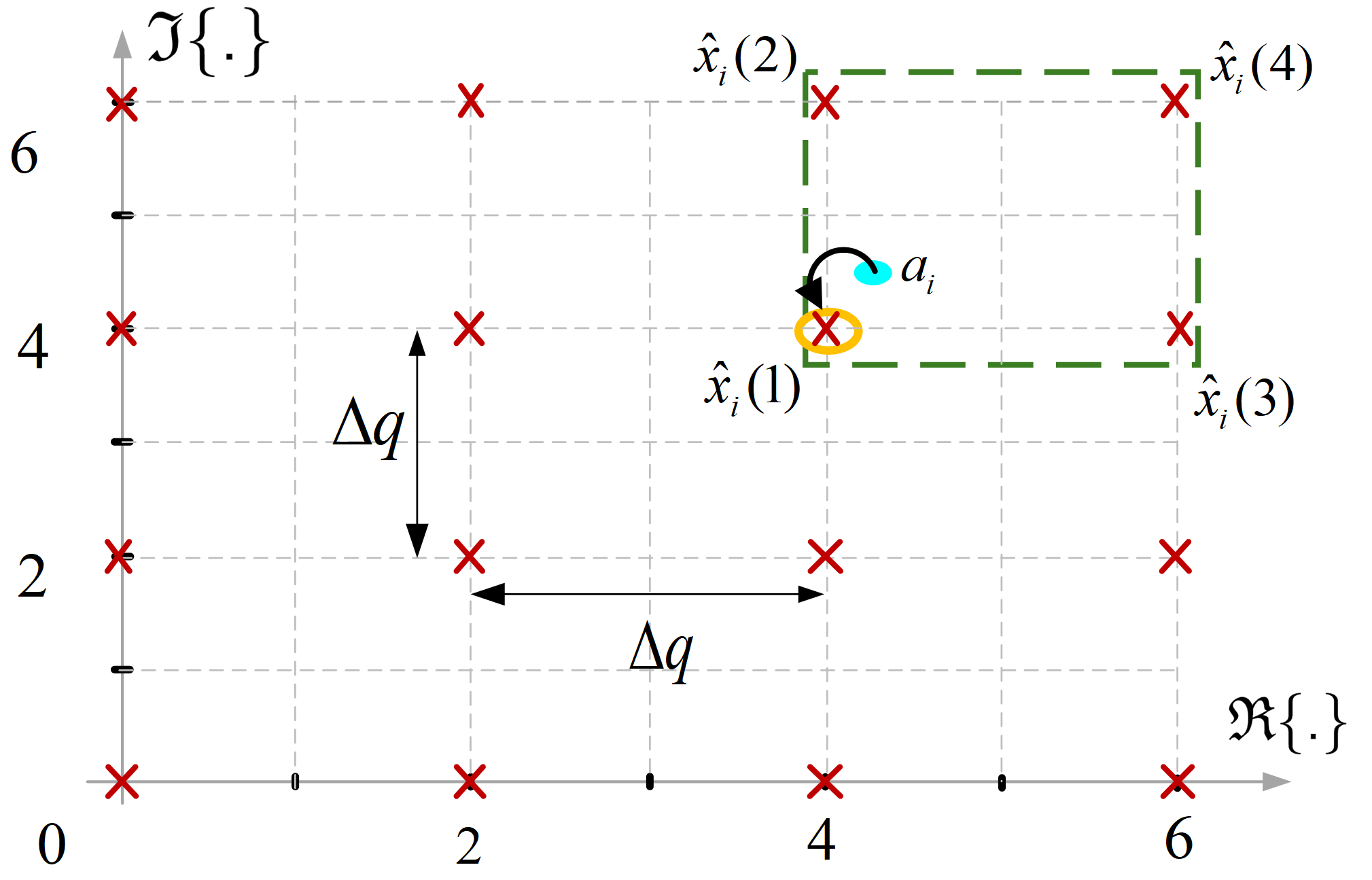} 
        \caption{Box decoding example with box size $\mathcal{B}=4$ (\cite{box}). }
        \label{fig:QAM_Box_diagram}
        \vskip -1\baselineskip plus -1fil
\end{figure}

\subsection{Box Decoding Algorithm}
\label{sec:Box}
Box Decoding starts at layer $i=N$ by first obtaining the reference point (also the ZF estimate)
$a_N=\frac{\tilde{y}_N}{r_{N,N}}$. This is followed by quantizing the reference point to the QAM constellation point in the lower-left corner via $\left\lfloor\frac{\tilde{y}_N}{r_{N,N}}\right\rfloor$, as shown in yellow in Fig.~\ref{fig:QAM_Box_diagram}. For any subsequent layer further down the tree, the reference point is computed as $a_i=\frac{\tilde{y}_i^{'}}{r_{i,i}}$, where $\tilde{y}_i^{'}=\tilde{y}_i-\sum_{j=i+1}^{N}r_{i,j}\hat{x}_j$ is the interference-canceled signal from the upper layers.\par 

Given a QAM spacing $\Delta q$, the \emph{Box Operation} constructs a box of size $\mathcal{B}$ around the reference point $a_i$ yielding $\{\hat{x}_i(l)\}^{\mathcal{B}}_{l=1}$ such that each $\hat{x}_i(l)$ satisfies $\hat{x}_i(l)=\hat{x}^{\mathrm{re}}(n)+j\hat{x}^{\mathrm{im}}(m)$, with
\begin{equation}
\begin{aligned}
\hat{x}^{\mathrm{re}}(n) &=
\left(\left\lfloor\frac{\mathfrak{R}\{a_{i}\} - M}{\Delta q}\right\rfloor + n \right)\Delta q, \\
\hat{x}^{\mathrm{im}}(m) &=
\left(\left\lfloor\frac{\mathfrak{I}\{a_{i}\} - M}{\Delta q}\right\rfloor + m \right)\Delta q,
\end{aligned}
\label{eqn:box_operation}
\end{equation}
where $n,m \in \left[-\frac{\sqrt{\mathcal{B}}}{2}+1,\frac{\sqrt{\mathcal{B}}}{2}\right]$ and $M = \sqrt{A} - 1$ defines the maximum constellation index per dimension.

It then repeatedly applies the above quantization and Box Operation as it traverses down to $i=2$. In the absence of pruning, each $i$-th layer leads to a total of $\mathcal{B}^{N-i+1}$ \textit{candidate nodes} corresponding to each $a_i$. At the bottom layer ($i=1$), it outputs the path $\{\hat{x}_N,\hat{x}_{N-1},\dots,\hat{x}_1\}$ that minimizes the total accumulated PED value across all layers.  To address the exponentially growing candidate nodes, pruning can be introduced at designated layers. Based on the chosen rule, only a subset of child nodes is retained and expanded further. The proposed pruning strategies are described in detail in the following section.


\section{Low-Complexity Box Candidate Pruning}
\label{sec:pruning_strategies}
\subsection{Efficient PED Comparison Techniques}
Each parent node at layer $i+1$ generates a fixed set of candidate nodes, referred to as a \textit{box cluster} at layer $i$. Since all candidates in a cluster originate from the same parent and differ only in their positions within the QAM grid, we can exploit the regular structure of the QAM grid to derive low-complexity approaches for comparing the PEDs. 

\subsubsection{\textbf{Local PED Minimum Rule (Rule~1)}}
Assuming a box size of \( \mathcal{B}=4 \) at layer $i$, the PED associated with each node \(\hat{x}_{i}(l)\) can be derived from \eqref{PEDa} as
\begin{equation}
d_{i,l} = d_{i+1} + r_{i,i}^2 | a_i - \hat{x}_{i}(l) |^2; \quad l = 1,\dots,\mathcal{B},
\label{PEDoneclu}
\end{equation}
where $d_{i+1}$ is the accumulated PED from upper layers. 

Since \(d_{i+1}\), \( a_i \) and \( r_{i,i} \) are identical for all candidates within a box cluster, their PED differences arise solely from the Euclidean distance term. For instance, the PED difference between two candidates \(\hat{x}_{i}(2)\) and \(\hat{x}_{i}(1)\) is given by
\begin{align}
\label{PEDdiff21}
d_{i,2} - d_{i,1} &\overset{(\mathrm{a})}{=} r_{i,i}^2 \left( | a_i - \hat{x}_{i}(2) |^2 - | a_i - \hat{x}_{i}(1) |^2 \right)\\ \notag
&\overset{(\mathrm{b})}{=} -r_{i,i}^2 \left( 2\mathfrak{I}\{\delta_1\} - \Delta q \right) \Delta q,
\end{align}
where \(\delta_1 = a_i - \hat{x}_{i}(1)\). Step (a) follows from the structure of the PED expression, while (b) assumes that the two candidates differ only in their imaginary components, i.e., \(\mathfrak{I}\{\hat{x}_{i}(2)\} = \mathfrak{I}\{\hat{x}_{i}(1)\} + \Delta q\), with identical real components, as illustrated in Fig. \ref{fig:QAM_Box_diagram}. This leads to the simple comparison rule:
\begin{equation}
\Delta q - 2\mathfrak{I}\{\delta_1\} \lesseqgtr 0,
\label{PEDdiff21rela}
\end{equation}
which allows us to determine whether \(\hat{x}_{i}(1)\) or \(\hat{x}_{i}(2)\) yields a smaller PED \textit{without explicitly performing distance calculations}. A similar rule applies when comparing candidates that differ along the real axis, leading to the corresponding decision rule:
\begin{equation}
\Delta q-2\mathfrak{R}\{\delta_1\}\lesseqgtr0.
\label{PEDdiff31rela}
\end{equation}
Due to the geometric symmetry of the QAM constellation, several PED differences among candidates within a box cluster are equal. For example, the PED difference between \(\hat{x}_{i}(2)\) and \(\hat{x}_{i}(1)\) is equal to that between \(\hat{x}_{i}(4)\) and \(\hat{x}_{i}(3)\); similarly, the difference between \(\hat{x}_{i}(3)\) and \(\hat{x}_{i}(1)\) matches that between \(\hat{x}_{i}(4)\) and \(\hat{x}_{i}(2)\). These relationships can be expressed as:
\begin{equation}
\begin{aligned}
d_{i,3} - d_{i,1} = d_{i,4} - d_{i,2};\quad
d_{i,2} - d_{i,1} = d_{i,4} - d_{i,3}.
\end{aligned}
\label{PEDdiffequal}
\end{equation}
This symmetry enables us to identify the best candidate (i.e., the one with the minimum PED increment) within a box cluster using only the simplified comparisons in \eqref{PEDdiff21rela} and \eqref{PEDdiff31rela}, collectively referred to as \textit{Rule~1}. \par
From a computational cost perspective, the PED comparisons in \textit{Rule~1} involve constant left shifts (multiplication by 2), which are typically implemented as hardwired bit-shifts and require negligible logic. Furthermore, the offset \(\delta_1 = a_i - \hat{x}_{i,1}\), required in \textit{Rule~1}, is already computed during the Box Operation step in candidate nodes generation.

\subsubsection{\textbf{Local PED Ordering Rule (Rule~2)}}
While \textit{Rule~1} identifies only the candidates with the maximum and minimum PED values within a box cluster, additional rules are required to determine the relative ordering of all candidates based on their PEDs. Extending the logic used in \eqref{PEDdiff21}, the PED differences between \(\hat{x}_{i}(3)\) and \(\hat{x}_{i}(2)\), as well as between \(\hat{x}_{i}(4)\) and \(\hat{x}_{i}(1)\), lead to the following comparison rules:
\begin{equation}
\begin{aligned}
d_{i,3} -d_{i,2} \lesseqgtr0 &\Rightarrow \mathfrak{R}\{\delta_1\}-\mathfrak{I}\{\delta_1\} &\lesseqgtr0 \ \\
d_{i,4} -d_{i,1} \lesseqgtr0 &\Rightarrow \mathfrak{R}\{\delta_1\}+\mathfrak{I}\{\delta_1\} -\Delta q &\lesseqgtr0 .
\end{aligned}
\label{PEDdiff3241rela}
\end{equation}
Equations \eqref{PEDdiff21rela}, \eqref{PEDdiff31rela} and \eqref{PEDdiff3241rela} are collectively referred to as the \textit{Rule~2}, which determines the order in terms of the PED values of candidate nodes within a box cluster of size \( \mathcal{B}=4\). For larger box sizes (e.g., \( \mathcal{B}=16, 64 \)), \textit{Rule~2} is typically applied to the four nearest neighboring constellation points around \( a_i \), which is generally sufficient for candidate pruning in most practical systems.
 \begin{figure}[t]
        \centering 
        \includegraphics[width=0.45\textwidth]{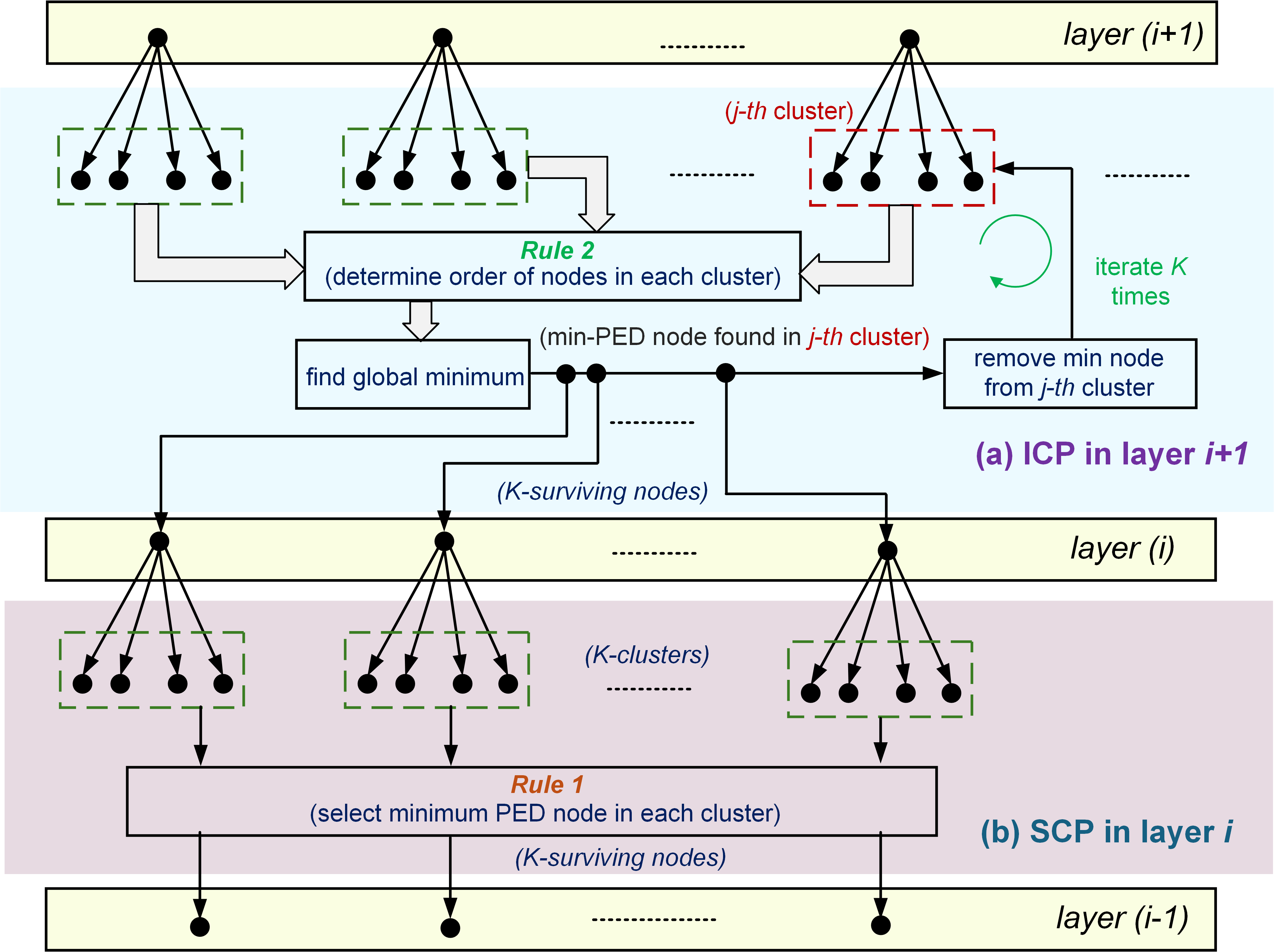} 
        \caption{ Block diagram of SICP pruning strategy combining a (a) Box-ICP in layer $i+1$ and (b) Box-SCP in layer $i$.}
        \label{fig:QAM_Box_SCP_ICP}
        \vskip -1.2\baselineskip plus -1fil
\end{figure}

\subsection{Proposed Candidate Pruning Strategies}
\subsubsection{\textbf{Single-Step Candidate Pruning (SCP)}}
The SCP strategy, as illustrated in Fig.~\ref{fig:QAM_Box_SCP_ICP}(b), selects one candidate node per box cluster at each detection layer using \textit{Rule~1}. Since candidates within a cluster share the same parent node and differ only in their relative constellation positions, \textit{Rule~1} enables efficient local comparisons without explicit PED computation or sorting. This allows a single child node to be chosen per parent, forming the core of the SCP pruning strategy. \par 
While SCP reduces complexity, it does not guarantee that the globally best candidates are retained, as pruning is performed independently across clusters. In particular, some clusters may contain multiple candidates with lower PEDs than those selected in other clusters, a limitation we refer to as the \textit{optimal child node skip problem}. To address this, we introduce the ICP strategy, which iteratively refines candidate selection across clusters.
\subsubsection{\textbf{Iterative Candidate Pruning (ICP)}}
To overcome the limitation of SCP, the ICP strategy adopts the MultiWay Merge (m-Merge) \cite{multiway} technique to iteratively generate the required child nodes. The m-Merge algorithm efficiently combines several pre-sorted source lists into a single ordered sequence by repeatedly selecting the global minimum from the local minima of all lists and replacing it with the next-best element from the same list. This approach avoids exhaustive candidate evaluations while ensuring that only the most promising nodes are expanded. \par

In the context of Box Decoding, each box cluster is treated as an individual source list for m-Merge operation. Before merging, candidates in each box cluster are pre-ordered using \textit{Rule~2}. The local minima from all clusters are then compared to identify the global best candidate, which is removed and replaced by the next element from the same cluster. This process is repeated until $K$ child nodes are selected. As illustrated in Fig.~\ref{fig:QAM_Box_SCP_ICP}(a), ICP is applied at a designated detection layer after the Box Operation has expanded the candidate list. Unlike the DKB algorithm \cite{DKB_conf}, which integrates m-Merge with explicit PED calculations, ICP leverages \textit{Rule~2} to maintain QAM-size independence while eliminating unnecessary computations. Moreover, by visiting only the required subset of candidates instead of evaluating all possible nodes at each stage as in conventional $K$-Best detection, ICP achieves substantial complexity reduction.

\subsubsection{\textbf{Hybrid SCP-ICP Candidate Pruning (SICP)}} The ICP strategy is introduced to overcome the optimal child-node skip problem and guarantees the selection of globally best candidates. However, full ICP across all layers incurs additional iterations that are not always necessary, since accurate pruning in the early layers already leads to substantial performance improvements. To balance complexity and accuracy, we propose a hybrid scheme, termed SCP–ICP (SICP$_t$), as illustrated in Fig.~\ref{fig:QAM_Box_SCP_ICP}. The scheme applies ICP only in the first $t$ layers and reverts to SCP in the remaining layers. 


\begin{table*}[t]
\caption{Complexity Comparison of proposed Box-SCP, Box-ICP and Box-SICP$_t$ with $K-$Best and DKB.}
\begin{center}
\begin{adjustbox}{width=\textwidth}
\begin{tabular}{|c|c|c|c|c|c|c|}
\hline
\multirow{2}{*}{Op}& \multirow{2}{*}{FLOPs} &\multicolumn{5}{c|}{\textbf{Detection Algorithms}} \\ 
 \cline{3-7}
 & &\textbf{\textit{\(K \)-Box\cite{KBox} $^{\mathrm{a}}$ }}& \textbf{\textit{DKB\cite{DKB_conf}}}& \textbf{\textit{Box-ICP}}& \textbf{\textit{Box-SCP}} & \textbf{\textit{Box-SICP$_t$}}\\
\hline
\multirow{2}{*}{IC} & RMUL& \multicolumn{5}{c|}{$2KN(N-1)$}\\
\cline{2-7}
                    & RADD& \multicolumn{5}{c|}{$2KN(N-1)$}\\
\cline{1-7}
\multirow{2}{*}{PC} & RMUL& $4(K^2+K^3)(\frac{N}{2}-1)+8K$ & $4(N-2)(2K-1)+8K$ & $4(N-2)(2K-1)+4(\mathcal{B}+K)$ & $4N\mathcal{B}$  & $ 4t(2K-1)+4(N-t)\mathcal B$\\
\cline{2-7}
                    & RADD& $3(K^2+K^3)(\frac{N}{2}-1)+6K$  & $3(N-2)(2K-1)+6K$  & $3(N-2)(2K-1)+3(\mathcal{B}+K)$ & $3N\mathcal{B}$ & $ 3t(2K-1)+3(N-t)\mathcal B$\\

\cline{1-7}
\multirow{2}{*}{CLE} & RMUL&  $4K(N-1)+4$  & $8(N-2)(K-1)$ & \multicolumn{3}{c|}{$4K(N-1)+4$} \\
\cline{2-7}
                     & RADD&  $K\sqrt{K}(N-2)+K\sqrt{A}$  & $8(K-1)(N-2)+K\sqrt{A}(N-1)$  &$K(N-2)(\sqrt{\mathcal{B}}+5)+2(K+1)$ & $K(N-2) (\sqrt{\mathcal{B}}+2)+2(K+1)$ & $K(N-2) (\sqrt{\mathcal{B}}+2)+3tK+2(K+1)$\\

\cline{1-7}           CP       & RADD& $3(x+y)-3(K^2-K+1)\, ^{\mathrm{b}}$   & \multicolumn{2}{|c|}{$(K(N-2)+1)(K-1)$}  & $K-1$ & $(Kt+1)(K-1)$\\
\cline{1-7}
\hline
\multicolumn{6}{l}{$^{\mathrm{a}}$ with the assumption in \cite{KBox} $K=\mathcal B$} \\
\multicolumn{6}{l}{$^{\mathrm{b}}$ $x=(\log_2 K+1)(\frac{1}{4}K^3\log_2 K-K),\,  y=K^3(3\log_2 K+1)$}
\end{tabular}
\label{tabCC}
\end{adjustbox}
\end{center}

\end{table*}

\section{Complexity Analysis}
The complexity is measured in terms of the number of real multipliers (RMUL) and real adders (RADD), expressed as $\mathcal{O}(\text{OP})=(\text{\#RMUL},\, \text{\#RADD})$, which together represent the dominant floating-point arithmetic operations (FLOPs). The main sources of complexity typically arise from the following operations:

\subsubsection{\textbf{Interference Cancellation (IC)}} At each layer $i$, the interference of detected symbols in the upper layers ($>i$) is removed from the current received symbol $\tilde{y}_i$. This requires $4(N-i)$ RMULs and $4(N-i)$ RADDs per candidate, yielding $\mathcal{O}(\text{IC})=(4(N-i), 4(N-i))$.

\subsubsection{\textbf{PED Calculation (PC)}} The PED value is computed for visited nodes using (\ref{PEDa}), with a fixed per-candidate complexity of $\mathcal{O}(\text{PC})=(4,3)$. For $K-$Box, since sorting is performed every two layers, the number of visited nodes equals $(\frac{N}{2}-1)(\mathcal B^2 + \mathcal B^3)+2\mathcal B$ \cite{KBox}. In DKB, the top layer visits $K$ nodes, while each subsequent layer visits only $(2K - 1)$ nodes due to the on-demand expansion scheme \cite{DKB_conf}. In contrast, only $(2K - 1)$ and $\mathcal{B}$ nodes are visited per ICP and SCP layer, respectively. It is important to note that in Box-SCP, full accumulated PED comparisons across paths are required only in the final layer.

\subsubsection{\textbf{Candidate List Expansion (CLE)}} Before PED computation, candidate nodes must be generated from the constellation. For DKB, each layer with $K$ surviving paths ($1<i<N$) starts with a rounding operation that requires $K\sqrt{A}$ real additions (RADDs) to identify the first child node, followed by Schnorr-Euchner (SE) enumeration to generate the remaining candidates. This results in a per-layer CLE complexity of $(8(K-1),\,8(K-1)+K\sqrt{A})$ in terms of RMULs and RADDs.\par

For Box-based detectors, candidate generation is performed via the Box Operation in \eqref{eqn:box_operation}, which requires $4$ RMULs and $\sqrt{\mathcal{B}}$ RADDs per box. Additional pruning incurs negligible overhead: \textit{Rule-1} and \textit{Rule~2} require only $2$ and $5$ extra RADDs per box for SCP and ICP, respectively. Importantly, the CLE complexity of Box-based schemes is independent of the constellation size $A$ and depends only on the box size $\mathcal{B}$. The total CLE complexity across all layers for different detection schemes is summarized in Table~\ref{tabCC}.

\subsubsection{\textbf{Candidate Pruning (CP)}}
In the $K$-Box algorithm \cite{KBox}, candidate pruning is performed via the $P$-to-$Q$ odd--even merge max-set selection method from \cite{merge_max_set}, with a sorting complexity of $\mathcal{O}(P\log_2^2 Q)$. In contrast, SCP incurs no additional pruning overhead, as it selects one candidate per cluster directly according to \textit{Rule~1}, which is already accounted for in the CLE analysis. Both DKB and ICP employ the m-Merge algorithm to select $M$ candidates from $L$ pre-ordered lists, resulting in a pruning complexity of $M(L-1)$ RADDs. The total CP complexity depends on the pruning method and the number of pruning stages applied throughout the detection tree, and is summarized in Table~\ref{tabCC}.

\section{Simulation Results}
\label{SimSec}

\subsection{BER Performance}

This section evaluates the BER performance of the proposed Box-SCP, Box-ICP, and hybrid Box-SICP$_t$ schemes. The results are benchmarked against linear ZF and LMMSE, near-optimal SD, conventional $K$-Best ($K=4$), and standard Box Decoding without pruning. Since DKB differs from $K$-Best only in its candidate expansion and selection strategy, yielding identical BER for the same $K$, it is omitted from the plots. For fairness, the number of surviving nodes per layer is fixed at $K = \mathcal{B} = 4$ for all tree-search-based schemes. Simulations are performed over an i.i.d. Rayleigh flat-fading MIMO channel with perfect CSI at the receiver, consistent with \cite{DKB_conf, Box_Prob}.

\begin{figure}[t]
    \begin{subfigure}{0.23\textwidth} 
        \centering 
        \includegraphics[width=0.9\textwidth]{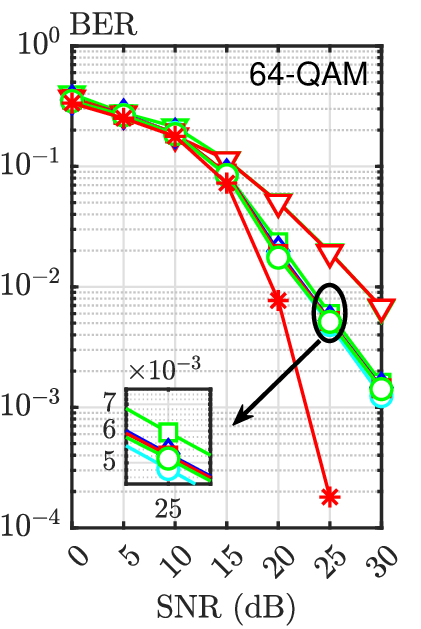}
        \caption{$4\times4$ MIMO} 
        \label{fig:4x4MIMO64QAM} 
    \end{subfigure}%
    \begin{subfigure}{0.23\textwidth} 
        \centering 
        \includegraphics[width=0.9\textwidth]{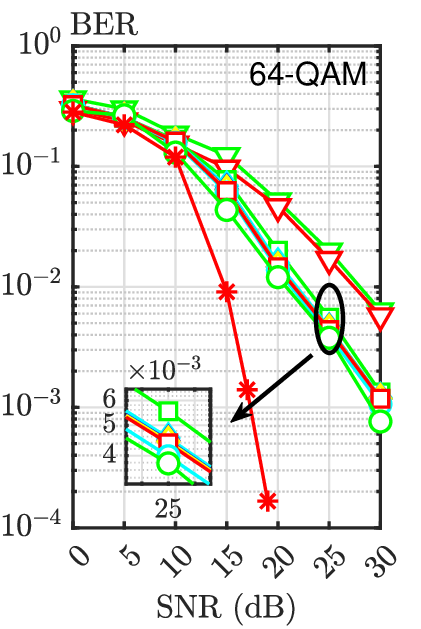} 
        \caption{$8\times8$ MIMO} 
        \label{fig:8x8MIMO64QAM} 
    \end{subfigure}
    
    \label{fig:BER64QAM} 

    \caption{BER performance of detection schemes (\textcolor{green}{$-\triangledown -$}: ZF; \textcolor{red}{$-\triangledown -$}: LMMSE; \textcolor{green}{$-\square -$}: Box-SCP; \textcolor{cyan}{$-\diamondsuit -$}: Box-SICP$_1$; \textcolor{blue}{$-\diamondsuit -$}: Box-SICP$_2$; \textcolor{yellow}{$-\diamondsuit -$}: Box-SICP$_3$;\textcolor{cyan}{$-\bigcirc -$}: $K$-Best; \textcolor{red}{$-\Box -$}: Box-ICP; \textcolor{green}{$-\bigcirc -$}: Box; \textcolor{red}{$-\ast -$}: SD.} 
    
    \vskip -1.25\baselineskip plus -1fil
\end{figure}

Fig.~\ref{fig:4x4MIMO64QAM} and Fig.~\ref{fig:8x8MIMO64QAM} show the BER performance for $4\times4$ and $8\times8$ MIMO systems with 64-QAM. For both antenna configurations, all Box variants achieve approximately 5 dB SNR gain over linear detectors and closely track $K$-Best performance. The effect of pruning is modest: Box-ICP exhibits negligible degradation relative to $K$-Best and Box Decoding without pruning, whereas Box-SCP incurs an SNR loss of about 1.3 dB. Introducing ICP layers mitigates this loss. Specifically, Box-SICP$_1$ improves performance by roughly 0.7 dB and approaches $K$-Best. Increasing $t>1$ yields negligible additional gain, as Box-SICP$_{2/3}$ performs almost identically to Box-SICP$_1$. This indicates that Box-SICP$_1$ provides an attractive performance–complexity trade-off\footnote{Due to space constraints, detailed BER curves are shown only for the $8\times8$ 64-QAM case. Similar trends are observed for larger MIMO dimensions and other modulation orders.}.

\subsection{Complexity and Performance Trade-off}

\begin{figure}[t]
    \begin{subfigure}{0.23\textwidth} 
        \centering 
        
        \includegraphics[width=\textwidth]{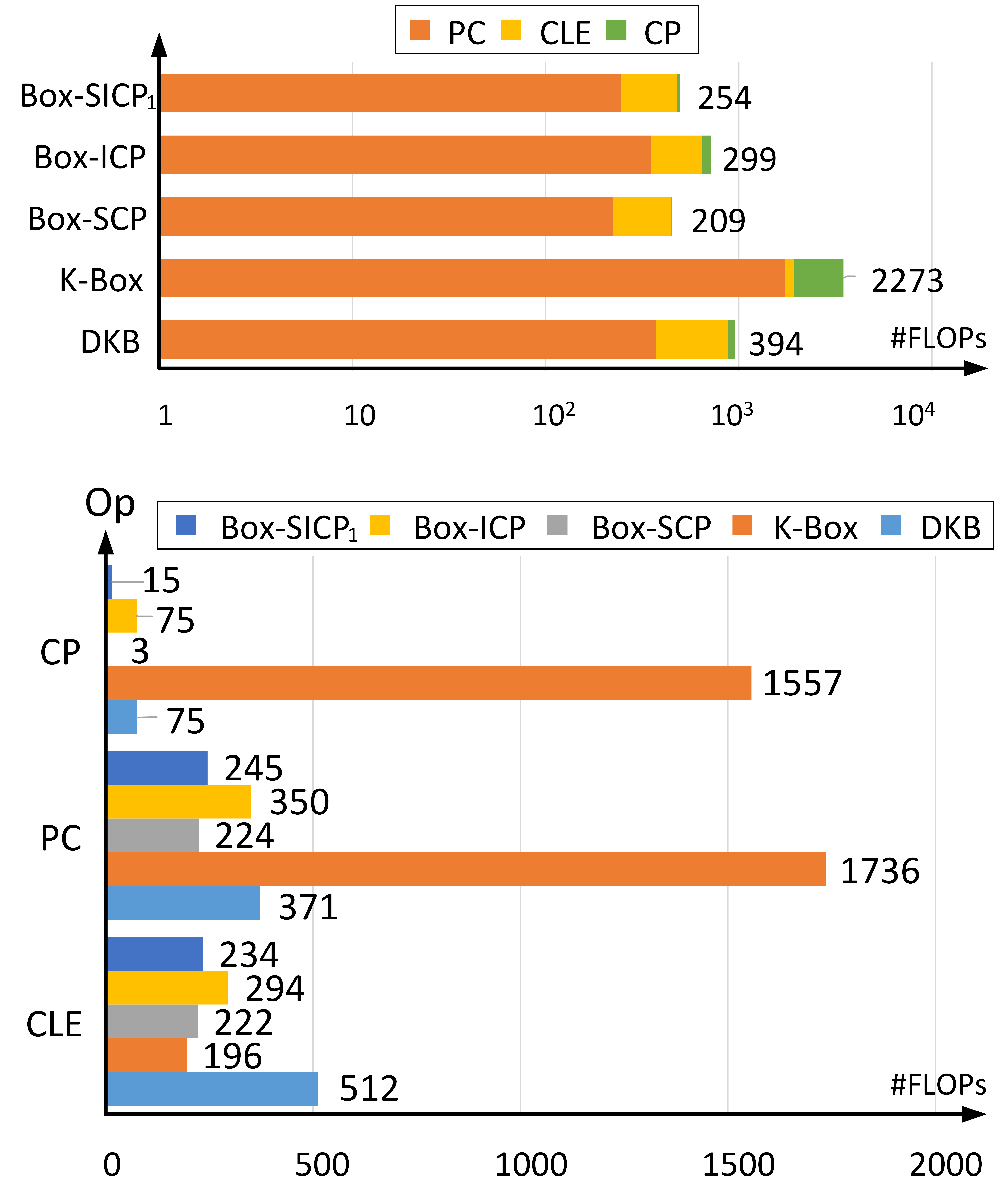}
        \caption{} 
        \label{fig:Flops_comp} 
    \end{subfigure}%
    \begin{subfigure}{0.23\textwidth} 
        \centering 
        \includegraphics[width=0.9\textwidth]{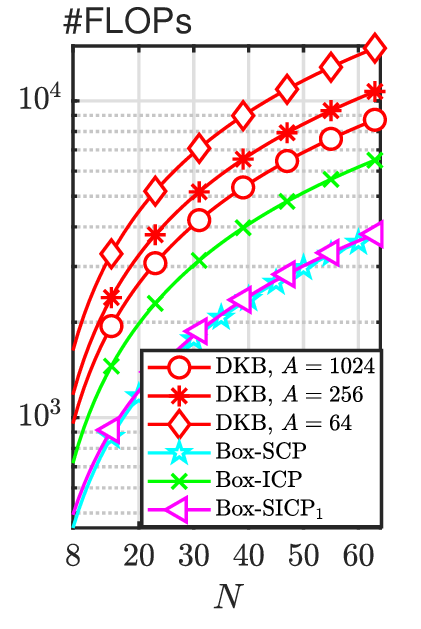}
        \caption{} 
        \label{fig:flpos_with_qam} 
    \end{subfigure}
    
    \label{fig:Complexity} 

   \caption{(a) Total complexity (top) and breakdown (bottom) for $8\times8$ MIMO with 64-QAM; (b) Total FLOPs as a function of MIMO dimension $N$ (logarithmic scale).}
    
    \vskip -1.2\baselineskip plus -1fil
\end{figure}

The overall complexity comparison for $8\times8$ MIMO with 64-QAM is shown in Fig.~\ref{fig:Flops_comp}. $K$-Box exhibits the highest complexity (2273 FLOPs), while DKB requires 394 FLOPs. The proposed schemes achieve further reduction. For example, Box-SCP (209 FLOPs) and Box-SICP$_1$ (254 FLOPs) reduce complexity by 47.0\% and 35.5\%, respectively, relative to DKB. Compared with sorting-based tree-search detectors (e.g., conventional $K$-Best and $K$-Box), the proposed strategies eliminate the area-critical sorting block. As shown in Fig.~\ref{fig:Flops_comp}, the sorting overhead in $K$-Box alone requires 1557 FLOPs. In contrast, the proposed sort-free approaches significantly reduce candidate pruning cost. Specifically, the m-Merge network in ICP introduces only 12 FLOPs per pruning layer, while SCP requires comparisons only in the final layer with a negligible cost of three FLOPs.\par

From a hardware perspective, the proposed methods also shorten the critical path. For example, $K$-Box implemented with a low-latency sorting network \cite{KBox,merge_max_set} requires 13 $T_{\text{RADD}}$ (adder cycles) to generate surviving nodes (assuming a $K^3$-to-$K$ max-set selection unit with $K=\mathcal{B}=4$), whereas fully parallel implementations of \textit{Rule~1} and \textit{Rule~2} enable SCP and ICP to produce survivors within one and three $T_{\text{RADD}}$, respectively.\par

Additionally, the savings become more pronounced for larger MIMO dimensions and higher-order QAM. For example, DKB locates the first child node using a slicing structure \cite{KBest_VLSI}, whose cost scales with QAM order $A$. In contrast, \textit{Rule~1} identifies the nearest constellation point using only two comparators, and \textit{Rule~2} orders the four closest points using five comparators, as reflected in the CLE breakdown in Fig.~\ref{fig:Flops_comp}. Box decoding inherently avoids QAM-dependent scaling due to its quantization-based candidate generation. The proposed pruning strategies preserve this property, as the operation counts of pruned Box variants (Table~\ref{tabCC}) do not scale with constellation size. Fig.~\ref{fig:flpos_with_qam} further illustrates this QAM-order independence and shows the scalability with respect to the MIMO dimension, where the proposed schemes maintain substantially lower complexity as $N$ increases.

\section{Conclusion}

This paper proposed \textit{sort-free} and \textit{QAM-order-independent}, deterministic low-complexity candidate pruning strategies, namely SCP, ICP, and their hybrid extension, to control the exponential node growth of Box Decoding in large-scale MIMO systems. By exploiting the geometric regularity of the QAM grid, the proposed methods enable deterministic-complexity candidate selection over SNR while preserving near–$K$-Best performance with substantial computational savings. In particular, Box-SCP supports feed-forward, \textit{fully parallel hardware implementation}, while Box-SICP$_1$ offers an improved performance–complexity trade-off for BER-critical scenarios. These results establish Box Decoding as a scalable and hardware-efficient solution for high-throughput MIMO detection. Future work will focus on detailed hardware architecture design and synthesis-level evaluation.

\bibliographystyle{IEEEtran}

\bibliographystyle{IEEEtran}

\end{document}